\documentclass[journal=ancham,manuscript=letter,layout=twocolumn]{achemso}

\usepackage{chemformula} 
\usepackage[T1]{fontenc} 

\author{Caroline Ceribeli}
\affiliation{S\~ao Carlos Institute of Chemistry, University of S\~ao Paulo, S\~ao Carlos, SP, Brazil}
\alsoaffiliation[Second University]{Department of Food Science, University of Copenhagen, Copenhagen, Denmark}
\email{caroline.ceribeli@usp.br}
\author{Henrique F. de Arruda}
\affiliation[Second University]{S\~ao Carlos Institute of Physics, University of S\~ao Paulo, S\~ao Carlos, SP, Brazil}
\author{Luciano da F. Costa}
\affiliation[Second University]{S\~ao Carlos Institute of Physics, University of S\~ao Paulo, S\~ao Carlos, SP, Brazil}

\title[]{How Coupled are Mass Spectrometry and Capillary Electrophoresis?}

\begin{document}
\normalsize







\begin{abstract}
The understanding of how science works can contribute to making scientific development more effective. In this paper, we report an analysis of the organization and interconnection between two important issues in chemistry, namely \emph{mass spectrometry} (MS) and \emph{capillary electrophoresis} (CE). For that purpose, we employed science of science techniques based on complex networks. More specifically, we considered a citation network in which the nodes and connections represent papers and citations, respectively. Interesting results were found, including a good separation between some clusters of articles devoted to instrumentation techniques and applications. However, the papers that describe CE-MS did not lead to a well-defined cluster. In order to better understand the organization of the citation network, we considered a multi-scale analysis, in which we used the information regarding sub-clusters. Firstly, we analyzed the sub-cluster of the first article devoted to the coupling between CE and MS, which was found to be a good representation of its sub-cluster. The second analysis was about the sub-cluster of a seminal paper known to be the first that dealt with proteins by using CE-MS. By considering the proposed methodologies, our paper paves the way for researchers working with both techniques, since it elucidates the knowledge organization and can therefore lead to better literature reviews.
\end{abstract}

\section{Introduction}
The understanding of how science works is fundamental to improve the way science is conducted~\cite{fortunato2018science}.  The area that deals with this type of analysis is called \emph{science of science}~\cite{fortunato2018science}.  Some examples of these studies include the analysis of the integration between theoretical and applied physics~\cite{de2018integrated}, the relationship between science and technology~\cite{meyer2010can}, and the analysis of subjects of photonic crystals~\cite{sinha2015overview}, among many others. In the chemistry field, some remarkable results have been reported,  such as the comparison of chemical paper citations from different datasets~\cite{bornmann2009convergent}, the analysis of the most cited books of analytical chemistry~\cite{braun2001peer}, and the possible relationship between quality and the number of citations of chemistry articles~\cite{bornmann2012factors}. Researchers have also considered simple statistics regarding the papers published in a given area in order to assist literature reviews~\cite{schmitt2003capillary}.

Analytical chemistry is an important sub-area of chemistry, and its several techniques are employed in many applications. Here, we focused on the analysis of the relationship between two essential techniques, which are capillary electrophoresis (CE) and mass spectrometry (MS).  The first consists of a separation process based on the movement of electrically charged particles or molecules in a conductive medium (in general liquid) under the influence of an electric field~\cite{baker1995capillary}. Then, the separation depends on the charge-to-size ratios of the analytes. MS is a technique of analysis based on the conversion of atoms and molecules into gas-phase ions. This phenomenon occurs due to the insertion of charge in a given analyte of interest. The trajectories of these ions are measured under the influence of electromagnetic fields, and the separation occurs as a consequence of their mass-to-charge ratios~\cite{pavia2014introduction}.  

CE and MS are essential techniques in analytical chemistry when used coupled or separately. The pioneering study that considers CE coupled with MS (CE-MS) was proposed by~\emph{Olivares et al.}~\cite{olivares1987line}, in which the authors employed an electrospray ionization (ESI) source. In this mode, a spray of charged droplets is produced. The solvent evaporates, causing shrinkage of charged droplets and disintegration of the drops resulting in the formation of gas-phase ions. ESI is the most common type of molecules ionization used for CE-MS~\cite{schmitt2003capillary,haselberg2007capillary}.

Some applications involving CE-MS include forensic~\cite{schmitt2003capillary,wey2001head}, environment~\cite{rodriguez2003off}, pharmaceutics~\cite{haselberg2013low}, and mainly, bio-omics~\cite{robledo2014review}. In the latter, the vast majority of studies regards to proteomic analysis~\cite{haselberg2007capillary, haselberg2013low}, but also includes other applications, such as metabolomics~\cite{soga2003quantitative, monton2007metabolome}, and genomics~\cite{mehrotra2013evaluation}. In this context, CE-MS techniques are employed in the analysis of biomarkers of body fluids related to diseases~\cite{zurbig2006biomarker}, analysis of quality markers in foods~\cite{perez2019advances}, studies of affinity interactions~\cite{nevidalova2019capillary}, among many others. 

CE has gained visibility over traditional separation techniques such as Liquid Chromatography (LC) due to the CE advantages, which include higher efficiency resulting in better resolutions of analytes, selective and faster separations, versatility, and smaller amounts of samples and reagents with less waste disposal~\cite{nevidalova2019capillary, robledo2014review}. Furthermore, the coupling with MS instead of the Ultraviolet-Visible (UV-Vis) spectrometer, which is one of the most common of the CE detectors, provides higher detection limits, identification, and structural information from the analyte molecule~\cite{nevidalova2019capillary}. Because of these requirements, CE-MS became a powerful technique of analyses in so many areas of chemistry to different purposes.

Because of the significant number of published papers involving both these areas, as well as the intricate relationship between them, we modeled the analyzed data by using tools developed in the network science area~\cite{costa2011analyzing}. The relationships between papers is modeled as a network, in which articles represent nodes that are connected according to their citations~\cite{fortunato2018science}. This type of approach can contribute to better understanding the analyzed areas. Interesting results have been obtained, including well-defined clusters in the citation network. Their respective key-words were found to define well-known subjects regarding the studied area. Also, we employed a multi-scale analysis that was able to provide a better understanding of seminal papers.

\begin{figure*}[!h]
  \includegraphics[width=0.86\textwidth]{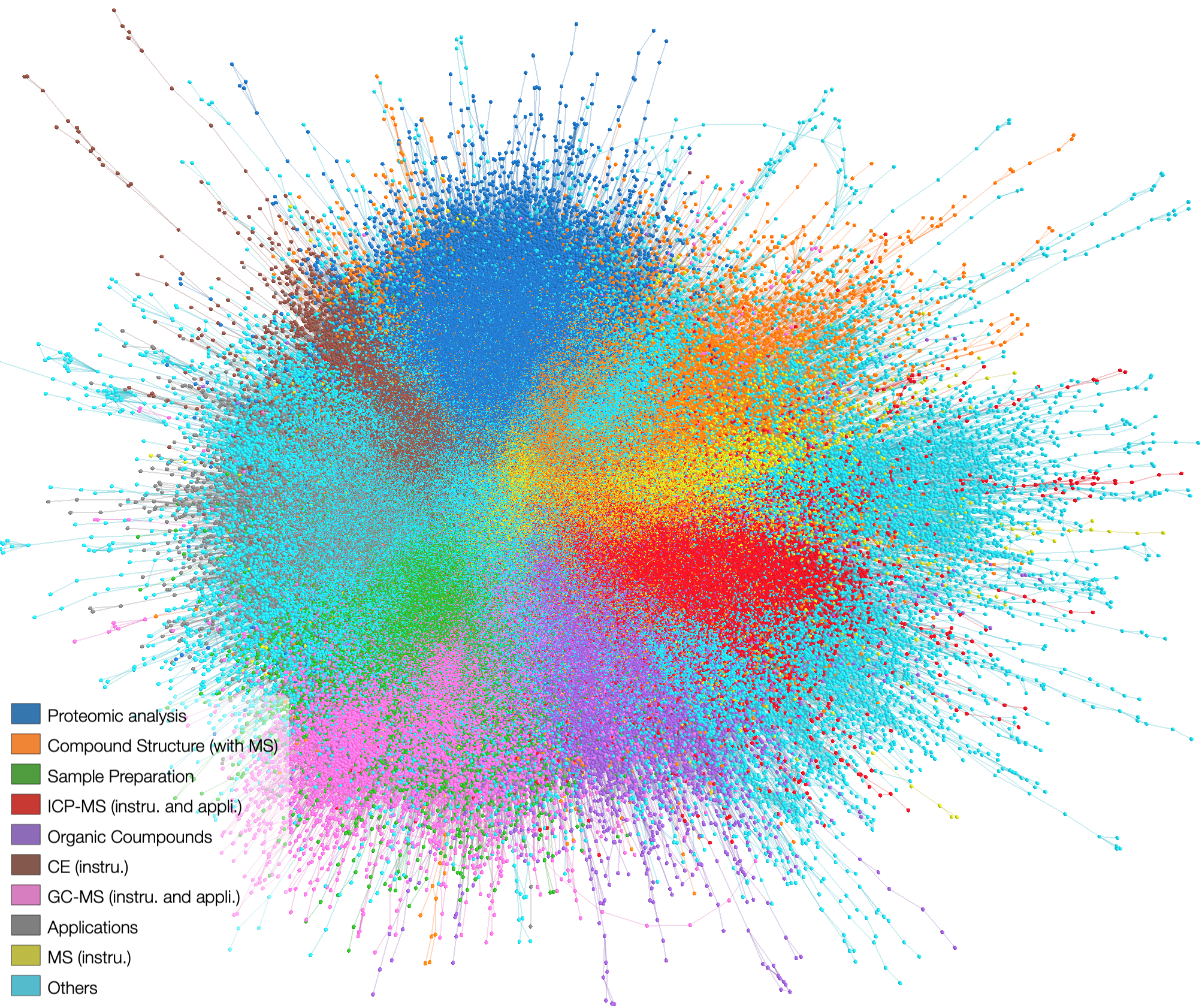}
  \caption{Visualization of the citation networks that comprises both areas, CE and MS, and the organization of the clusters with the respective detected subjects. The named clusters represent the nine groups with the highest number of papers, and \emph{others} comprises the set of all the remaining papers. The legend of the figure lists the cluster names in decreasing order according to the number of papers. The employed visualizer was developed by \emph{Silva et al.}\cite{silva2016using}.}
  \label{fgr:network}
\end{figure*}

\section{Materials and Methods}
Here, we analyze a large set of articles obtained from the \emph{Microsoft Academic Graph} (MAG)~\cite{sinha2015overview}, which contains information of the paper titles, their respective citations, and their abstracts, among other fields. More details regarding the dataset are provided in the MAG's web page\footnote{\url{https://www.microsoft.com/en-us/research/project/microsoft-academic-graph/}}. The employed data set was obtained in 2018. In order to filter the desired papers, we search for some key-words in the paper's abstracts, as follows: \emph{capillary electrophoresis}, \emph{mass spectrometry}, and \emph{electrospray ionization}.  The obtained network is shown in Figure~\ref{fgr:network}. In order to remove non-desired papers, we considered only the largest connected component of the network.

In order to detect the main subjects of the network, we employed the method proposed by \emph{Silva et al.}~\cite{silva2016using}, which detects key-words from groups of nodes. The first step of this method consists of finding clusters of papers, also known as communities, which are defined as groups of nodes well-connected between themselves and weakly connected with the remaining of the network. Here, we detect clusters of articles by using the \emph{Infomap} method~\cite{rosvall2009map}, which has been used in related applications~\cite{rosvall2008maps,rosvall2010mapping}. This approach takes into account the information flow between the network nodes to find the clusters. Furthermore, we also considered an Infomap version that organizes these clusters in multi-scale fashion. In Figure~\ref{fgr:network}, we show the obtained organization on the larger scale. Details about the methods of automated taxonomies and clustering are provided in Supporting Information~S1.

\section{Results and discussion}
\begin{figure}[!h]
  \includegraphics[width=0.48\textwidth]{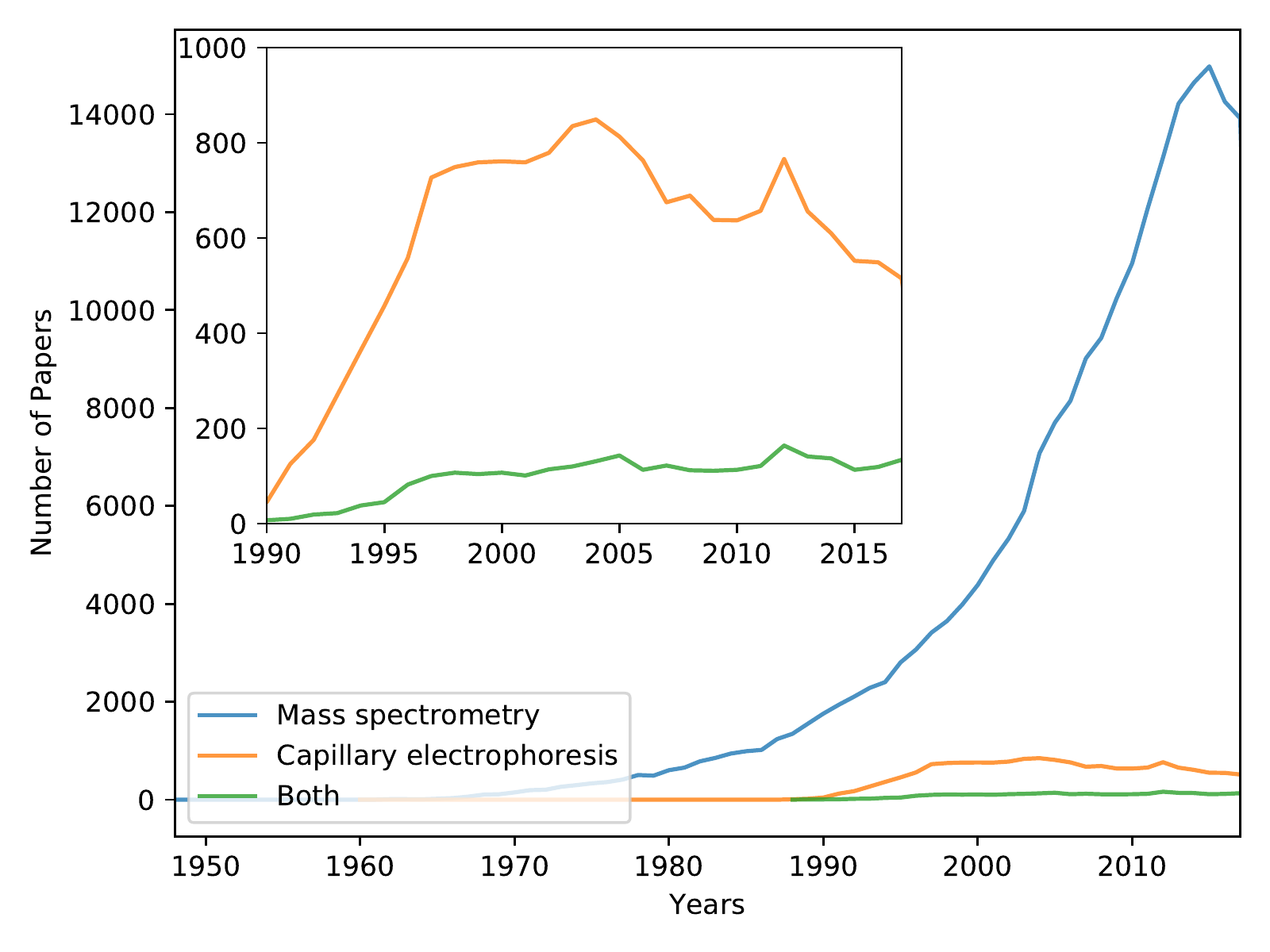}
  \caption{Number of papers for respective key-words along the years. The inset shows a zoomed region of the plot.}
  \label{fgr:temporal}
\end{figure}

First, we present some essential information regarding our network representation, which comprises 239.169 articles of both of the analyzed sub-areas (CE and MS). The number of links (citations) between these articles is 1.455.454. The number of papers along the years is shown in Figure~\ref{fgr:temporal}. CE-MS, CE, and MS represent 1.2\%, 5.9\%, and 92.9\% of the papers, respectively. The higher number of MS papers is found because this technique is more general and started to be successfully employed earlier than CE~\cite{schmitt2003capillary}. Additionally, MS is widely applied in analytical chemistry (with or without using coupled equipment). CE-MS is a relatively new technique~\cite{olivares1987line}, initiated approximately 30 years ago. This fact may account for the lack of a well-defined cluster entitled CE-MS (see Figure~\ref{fgr:network}). In the future, as these coupled techniques become more widespread, a more well-defined cluster may be obtained.

By employing the proposed methodology, interesting results were found. First, to identify the computed clusters, we considered their respective five key-words. For the sake of simplicity, we analyzed only the nine largest clusters. The obtained names are shown in Figure~\ref{fgr:network}. For instance, the first group is called \emph{Proteomic analysis} and the respective selected key-words ordered by importance are: i - ``protein'',  ii - ``identify'',  iii - ``cell'', iv - ``peptide'',  and v - ``proteomic''. The complete list of words that gave rise to the subject names are shown in Supporting Information~S2. 

We found that the vast majority of the articles devoted to CE are organized inside the same cluster dedicated to the instrumentation of CE, called \emph{CE (Instru.)}, which contains 14,089 papers. Additionally, 73.8\% of the papers' abstracts of \emph{CE (Instru.)} were found to include the word ``electrophoresis'' and do not have the word ``mass''.  This result indicates that our proposed methodology is reflecting the information of the organization of the subjects. Furthermore, in this cluster, only 7.5\% of the abstracts have both expressions (``mass'' and ``electrophoresis''). The remaining of the papers devoted to CE-MS were found to be spread among all of the network clusters. For more details, see Supporting Information~S3. Both techniques are essential to the development of analytical chemistry when used coupled or independently. When these techniques are coupled, the mass spectrometer works as detector and analyzer. In contrast, other types of detectors can be employed together CE, including UV absorbance, fluorescence, laser-induced fluorescence, conductivity, amperometric, radiometric, and refractive index~\cite{baker1995capillary}.

By considering CE-MS, we can classify the use of these techniques as being of interest to instrumentation or applications. Regarding the instrumentation aspects, we take into account the organization and type of equipment; the variables that influence the given analysis; and the optimization of the conditions. For example, \emph{CE (instru.)} involves decisions about the used detector, buffer solution, and the specification of the capillary, etc. On the other hand, the focus on applications mainly considers CE-MS as a tool, with  emphasis on the sample that is analyzed and the advantages that these coupled techniques can provide.

In the case of MS, the instrumentation is related to the employed ionization source and the type of mass analyzer, among other configurations. These articles are represented by the cluster called \emph{MS (Instru.)} (see Figure~\ref{fgr:network}). However, the majority of the articles were found to be devoted not to instrumentation, but applications. 

The largest of the analyzed clusters is \emph{proteomic analysis}, which includes all of the investigated techniques (CE, MS, and CE-MS). This cluster comprises 15.0\% of the entire network, and 8.8\% of the articles of CE-MS are found in this cluster. This technique is particularly interesting for protein analysis because it allows the study of intact proteins~\cite{haselberg2013low,haselberg2007capillary}. Before the advent of CE-MS, the more commonly employed methods to characterize proteins in proteomic approach were LC-MS and MS/MS. However, in these cases, proteins are typically digested before the analysis. So, information regarding the secondary and tertiary structures of them are lost.

A significant number of papers were also found to be devoted to other applications, which are represented by \emph{compound structure (with MS)}, \emph{sample preparation}, \emph{organic compounds}, and \emph{applications}. However, the groups named \emph{ICP-MS (instru. and appl)} and \emph{GC-MS (instru. and appl)} incorporate both the articles of instrumentation and applications, where ICP and GC mean \emph{inductively coupled plasma} and \emph{gas chromatography}, respectively. Interestingly, in contrast with CE-MS, both GC-MS and ICP-MS articles were found in cohesive communities.

In order to better understand the organization of the obtained network, we selected one seminal paper that gave rise to the studies of the coupling between both techniques. This paper was written by Olivares et al.~\cite{olivares1987line} to analyze and separate a mixture with quaternary ammonium salts by using CE-MS with ESI. Interestingly, the primary focus of this paper is to propose the coupling technique, which agrees with the cluster classification, \emph{CE (Instru.)}. By exploring multi-scale analysis, in the next level of granularity, this paper was classified in a cluster with the following key-words: ``capillary electrophoresis'', ``separation'', ``electrospray'', ``charge'', and ``solution''. These key-words well represent the subjects of Olivares' paper. So, this article is substantially coherent with its subcluster, which reinforces its relevance in the area. \emph{Smith et al.}~\cite{smith1988capillary} published another study that complemented Olivares' seminal paper and performed an analysis of polipeptides and quaternary ammonium salts. As expected, both articles were found to be part of the same cluster and subcluster. 

To complement our study, we analyzed another seminal paper that paved the way for the analysis of proteins through CE-MS~\cite{loo1989peptide}. More specifically, the authors analyzed polypeptides and proteins (e.g., myoglobin). Counterintuitively, this paper was not found to be part of the \emph{proteomic} cluster, appearing instead in the \emph{Compound Structure (with MS)}. Probably this article is classified in this cluster because it also describes applications only employing MS. Since this paper is devoted to more than one objective, it can be part of this cluster because of the information flow in the network. More specifically, information flow is determined by the citations between articles. By considering the articles' subcluster, the computed key-words are: ``spectrum'', ``multiply charge'', ``dissociation'', ``protein'', and ``molecular ion''. In this case, ``protein'' is one of the key-words, which indicates that other articles of this subcluster can also be related to the analysis of proteins. 

\section{Conclusions}
In the present paper, we show the interrelationship between two sub-areas of analytical chemistry through an analysis based on the science of science and network science techniques. More specifically, we employ a methodology of cluster detection, considering the information flow of the network. By using these clusters, we detected the cluster key-words and determined their respective names. First of all, our results revealed that coupling between both techniques is not clustered in the network. So, the papers are more well-grouped according to their respective applications than to the employed analytical technique. A similar result was found regarding the relationship between biological and physical sciences~\cite{burke2019interdisciplinary}. However, for other coupling techniques, GC-MS and ICP-MS, the papers were found to be much more well-grouped in the citation network. In the first case, this effect can be explained by the age of these couplings, which are significantly older than CE-MS~\cite{gohlke1993early}. In the second case, although ICP-MS has approximately the same age as CE-MS technique~\cite{taylor2001inductively}, the result can be explained because ICP-MS is more widespread than CE-MS.

We also showed that the majority of the papers regarding CE belong to the cluster named \emph{CE (instr.)}, which indicates that a significant number of CE articles include words related to instrumentation. However, for MS articles, the most significant clusters are devoted to applications. So, it would be interesting for researchers to take into account these characteristics before performing a literature review. In contrast with these cluster patterns, the literature review can be much more challenging for CE-MS.  \emph{Schmitt-Kopplin et al.}\cite{schmitt2003capillary} already explored some simple statistics and \emph{Braun et. al.}~\cite{braun2001peer} showed the importance of didactic books, but automatic systems can complement the characterization.

In order to better understand the obtained cluster organizations, we employed a multi-scale analysis of seminal papers. The first analyzed article proposed the coupling between CE and MS, where the main cluster is related to CE, and the subcluster key-words are intrinsically related to the subject of this seminal paper. Additionally, we analyzed another seminal paper that proposed the first analysis of proteins with CE-MS. In this case, the main cluster was \emph{Compound Structure (with MS)}, but not \emph{proteomic analysis}. However, ``protein'' is one of its subcluster key-words. Therefore, both of the granularity bring essential information regarding this article, which can happen to other articles.

As future works, many different areas of study can be approached, including relationship between other sub-fields of chemistry. To obtain new insights regarding the most related papers in the citation network, the multi-scale analysis can also be employed to other specific papers. Furthermore, the presented analysis can assist literature reviews of a given desired area.

\begin{acknowledgement}
Caroline Ceribeli thanks Coordena{\c c}{\~a}o de Aperfei{\c c}oamento de Pessoal de N\'ivel Superior - Brasil (CAPES) - Finance Code 001. Henrique F. de Arruda acknowledges FAPESP for sponsorship (grant no. 2018/10489-0). Luciano da F. Costa thanks CNPq (grant no. 307085/2018-0) and NAP-PRP-USP for sponsorship. This work has been supported also by FAPESP grants 11/50761-2 and 2015/22308-2. Research carried out using the computational resources of the Center for Mathematical Sciences Applied to Industry (CeMEAI) funded by FAPESP (grant 2013/07375-0). The authors also thank Filipi Nascimento Silva for fruitful discussions and help concerning data.
\end{acknowledgement}

\newpage

\section{Supporting Information}

\subsection{S1 - Subject detection}
In order to detect clusters of articles and their respective subjects, we employ a variation of the method proposed by \emph{Silva et al.}\cite{silva2016using}. First of all, the network, $G$, is created from the dataset of papers and their respective citations. In the first part, we compute the $G$ communities. In the original paper, the authors employ the Fast Multilevel Method~\cite{blondel2008fast}, but here we use the Infomap~\cite{rosvall2009map}. The latter presents some advantages, such as the capability of finding smaller clusters. This method also considers the network directions and network flow and allows us to perform a multi-scale analysis. 

The detection of key-word start with the pre-processing of the papers' abstracts. Differently from the original paper, we considered only the abstracts. We pre-process these texts, as follows: (i) symbols and punctuation were removed, (ii) all the letters were converted into lower case, and (iii) the remainder of the words are lemmatized~\cite{manning1999foundations}. Furthermore, the key-words of the clusters are computed from $n$-grams of words. More specifically, $n$-grams are defined as sets of $n$ adjacent words. For example, in ``the relationship between mass spectrometry and capillary electrophoresis'', the 2-grams are: ``the relationship'', ``relationship between'', ``between mass'', ``mass spectrometry'', ``spectrometry and'', ``and capillary'', and ``capillary electrophoresis''. In the original paper, the authors considered $n$-grams with 1 and 2 words, but here we employ 1, 2, and 3.

In order to compute the key-words importance, the following steps are executed. First, by considering the communities, the normalized $n$-gram frequencies of in-cluster ($F^{in}_c$) is computed as
\begin{equation}
 F^{in}_c(w) = \frac{f_c(w)}{n_c},
\end{equation}
where, $f_c(w)$ and $n_c$ are the frequency of $w$ and the total number of $n$-grams in a given cluster $c$, respectively. The out-cluster ($F^{out}_c$) are computed as follows
\begin{equation}
 F^{out}_c(w) = \sum_{\alpha \neq c}{\frac{f_\alpha(w)}{N - n_c}},
\end{equation}
where $N$ is the number of network nodes and the frequencies accounts for all $\alpha$ communities excluding $c$. 

In the following, the importance index $I$ is computed as 
\begin{equation}
    I_c(w) = F^{in}_c(w) -F^{out}_c(w),
\end{equation}
where $c$ and $w$ are words and communities, respectively. This index is sorted and the first $N$ words are selected as the key-words of $c$. Due to the high number of articles, we consider $N=5$.

\subsection{S2 - Subjects and detected words}
By considering the subjects detected with the method presented in Section~S1, we manually assigned the cluster's names, as shown in Table~\ref{tab:key}. In this case, we considered only the five first key-words, which are ordered in decreasing order according to their importance.

\begin{table*}[!th]
\scriptsize
\begin{tabular}{|l|l|}
\hline
\textbf{Cluster Name}                           & \textbf{Detected key-words}                                                                                         \\ \hline
Proteomic Analysis                       & `protein', `identify', `cell', `peptide', and `proteomic                                                    \\ \hline
Compound Structure (with MS)             & `complex', `ion', `structure', `mass spectrometry', and `electrospray ionization'                           \\ \hline
Sample Preparation                       & `sample', `extraction', `limit', `liquid', and  `determination'                                              \\ \hline
ICP-MS (Instru. and Appli.) & `plasma mass spectrometry', `inductively couple plasma', `couple plasma mass', \\
 & `element', and `sample'     \\ \hline
Organic Coumpounds                       & `gas', `organic', `pyrolysis', `product', and `hydrocarbon'                                                 \\ \hline
CE instrumentation                       & `separation', `capillary electrophoresis ce', `detection', `buffer', and `ph'                               \\ \hline
GC-MS (Instru. and Appli.)  & `gas chromatography', `component', `gas chromatography-mass spectrometry',\\  
 & `essential oil', and `headspace' \\ \hline
Application                              & `drug', `liquid', `metabolite', `method', and `plasma'                                                      \\ \hline
MS instrumentation                       & `ion mass spectrometry', `secondary ion mass', `surface', `laser', and `time-of-flight'                     \\ \hline
\end{tabular}
\caption{Cluster names assigned by a specialist using the detected keywords.} \label{tab:key}
\end{table*}

\subsection{S3 - Visualization of the papers by key-words}
\begin{figure*}[!h]
\includegraphics[width=0.95\textwidth]{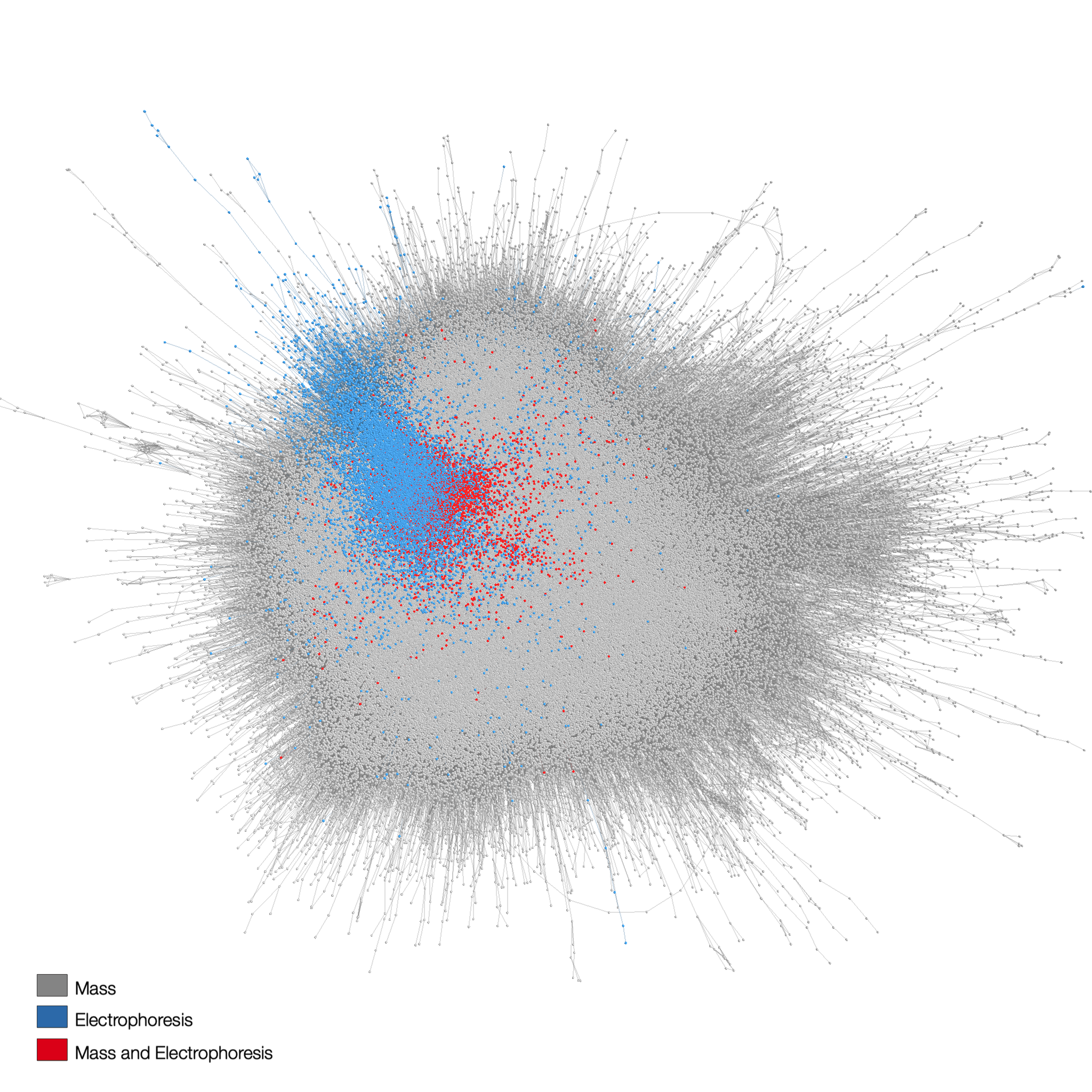}
\caption{Visualization of the citation networks with their respective key-words. Here, we highlighted the nodes that represent articles with the words "electrophoresis" and "mass and electrophoresis", which correspond to the papers of CE and CE-MS, respectively. The employed visualizer was developed by \emph{Silva et al.}\cite{silva2016using}.}
  \label{fgr:networkSupl}
\end{figure*}

In Figure~\ref{fgr:networkSupl}, we present the visualization of the network, in which we show the articles that represent the techniques. For that, we select the papers with the words "Mass" or "Electrophoresis" and with both of words, which define articles of \emph{mass spectrometry}, \emph{capilary electrophoresis}, and the use of both techniques, respectively. We use the word \emph{electrophoresis} instead of \emph{capillary electrophoresis} because the network regards only CE papers.

\bibliography{achemso-demo}
\end{document}